# Limitation of multi-resolution methods in community detection

Ju Xiang[1] and Ke Hu[2]

[1]*First Aeronautical College of Air Force, Xinyang 464000, Henan, China*
[2]*Department of Physics, Xiangtan University, Xiangtan 411105, Hunan, China*

E-mail: xiangju0208@yahoo.com.cn

**Abstract**
Recently, a type of multi-resolution methods in community detection was introduced, which can adjust the resolution of modularity by modifying the modularity function with tunable resolution parameters, such as those proposed by Arenas, Fernández and Gómez and by Reichardt and Bornholdt. In this paper, we show that these methods still have the intrinsic limitation—large communities may have been split before small communities become visible—because it is at the cost of the community stability that the enhancement of the modularity resolution is obtained. The theoretical results indicated that the limitation depends on the degree of interconnectedness of small communities and the difference between the sizes of small communities and of large communities, while independent of the size of the whole network. These findings have been confirmed in several example networks, where communities even are full-completed sub-graphs.

## 1. Introduction

Many complex networks, including social, biological and technological networks etc., consist of communities or modules—groups of vertices within which connections are dense while between which they are sparser [1-2]. Community detection is of considerable interest for analyzing the structure and function of the networks [2-4]. Recent years, many community-detection algorithms have been proposed based on various approaches [5-16]. Especially, the optimization of modularity—a quality function for community division of network which was proposed by Newman and Girvan—becomes the most popular strategy widely used in community detection [17-21].

For a given community division of network, the modularity function can be expressed as

$$Q = \sum_s \frac{k_s^{in}}{2M} - (\frac{k_s}{2M})^2, \quad (1.1)$$

where $M$ is the total number of edges in the network, $k_s^{in}$ and $k_s$ are respectively the inner degree and the total degree of group $s$, and the sum over all communities in the given network [22]. For weighted case, $M$ becomes the total weight of edges in the network, $k_s^{in}$ the inner weighted degree of the group $s$, $k_s$ the total weighted degree of group $s$ [23]. Generally, the larger the modularity, the better the division is. So, the modularity measure provides a way objectively to evaluate the quality of community divisions of networks, and it also suggests a kind of alternative strategy for community detection—the modularity optimization [17-22].

However, it has been proven that the modularity-optimization strategy has a resolution limit, failing to detect communities smaller than a scale that depends on the total number of edges in the network and the degree of interconnectedness of the communities [24]. That is to say, the small communities $s$ and $t$, even though they are full-connected sub-graphs connected only by single edges, will also be jointed together into a larger group by the modularity-optimization methods, when the total degrees $k_s$ and $k_t$ of them satisfy

$$k_s k_t < 2M \cdot E_{st} \quad (1.2)$$

where $E_{st}$ denotes the number of edges connecting communities $s$ and $t$, and $M$ is the total number of edges in the network. This is because the modularity is greater if the small communities are assigned into one group.

Recently, many schemes have been proposed to attack the resolution problem of modularity in community detection, such as, by introducing new quality function [25-26], by recursively partitioning sub-networks [27], by re-weighting the inter- and intra-edges in networks [28-30], and by directly or indirectly modifying the modularity function through tunable parameters [11, 31-32] (see ref [2] for reviews). In particular, a type of multi-resolution methods, which modifies the modularity function by tunable resolution parameters, can adjust the resolution of modularity by their parameters. For example, Reichardt and Bornholdt (RB) in ref [11] discussed a modified version of the modularity function which introduces a parameter to tune the contribution of the null model in the modularity, and Arenas, Fernández and Gómez (AFG) [32] also proposed a multi-resolution method by providing each vertex with a self-loop of the same magnitude $r$, which is equivalent to modifying the modularity function by the parameter $r$.

Both the two multi-resolution methods above can help to analyze the communities at different scales, by tuning their parameters. But these methods themselves, before being applied to the real problem of community detection, should have been thoroughly understood to ensure that the multi-scale structures found in networks are reasonable. The comparisons of these methods and the relations with other methods have been discussed in some papers [2, 32-33]. Here, we show that these methods have the intrinsic limitation themselves—with increasing the values of the parameters, large communities may have been split when small communities become gradually visible in some case where the communities even are full-connected sub-graphs. The reason is that it is at the cost of the community stability that the enhancement of the modularity resolution is obtained. And we show that the limitation depends on the degree of interconnectedness of small communities and the difference between the sizes of small communities and of large communities, while independent of the size of the whole network.

In the following sections, we present a critical analysis of the applicability of the multi-resolution methods to the problem of community detection. Firstly we study the multi-resolution process of the RB method and give the analytical formula of its limitation, by analyzing the expressions of parameters in four transition points where the whole network is regard as one large group, the (small) communities become visible, the (large) communities is to be split, and the whole network split into individual vertices. Secondly, the theoretical results are confirmed in several network examples. Then, we discuss the relationship between the RB method and the AFG method, and extend the limitation condition to the AFG method. Finally, we come to our conclusion.

## 2. Limitation of RB multi-resolution method
### 2.1 RB multi-resolution method

Reichardt and Bornholdt (RB) have discussed a kind of modified-version modularity function by tuning the contribution of the null model in the modularity with a parameter. The modified modularity function of RB [11, 32-34], referred to as RB modularity, can be written as

$$Q^{R\&B}(\gamma) = \sum_s \frac{k_s^{in}}{2M} - \gamma (\frac{k_s}{2M})^2 \tag{2.1}$$

where $\gamma$ is the pre-factor of tuning the contribution of the null model, other notations are the same as those in the Eq. (1.1). According to Eq.(2.1), the inequality (1.2) of the resolution becomes

$$k_s k_t < \frac{2M \cdot E_{st}}{\gamma}. \tag{2.2}$$

As we see, the RB method can adjust the resolution by the parameter $\gamma$.

The aims to which the RB method is designed may not be to study the resolution of modularity [11], while it indeed provides an alternative approach to the study of this issue. When $\gamma = 1$, the result obtained by optimizing the RB modularity is the same as that using the modularity $Q$ in Eq. (1.1). When $\gamma < 1$, we can find the superstructures above those at $\gamma = 1$, especially, if $\gamma \to 0$, then all vertices in network will be assigned into one large group. When $\gamma > 1$, we can have access to the substructures under those at $\gamma = 1$. When the value of the parameter is large enough, the whole network will be separated into a set of single-vertex groups by optimizing the RB modularity.

Moreover, according to Eq.(2.1), the contribution of each community $s$ to the modularity can be denoted by $q_s = k_s^{in}/(2M) - \gamma (k_s/(2M))^2$. The larger the value of $q_s$, the better the stability of the community, in terms of the Potts model where the energy corresponds to the negative of the modularity function in Eq.(2.1), i.e. $Q^{R\&B}(\gamma) = \sum_s q_s = -H(\gamma)/M$ [11]. When the resolution of modularity is improved by increasing the value of $\gamma$, the value of $q_s$ is to be decreasing. Thus the stability of the community will be weakened due to the increase of $\gamma$. Therefore, we can say, it is at the cost of the community stability that the enhancement of the modularity resolution is obtained.

**2.2 Analysis of multi-resolution process**

For the sake of clarity and to simplify the mathematical expressions (without affecting the final results), we analyze the multi-resolution process of the RB method through a kind of simple networks containing two types of well-defined communities with different sizes, where the large communities can be found by maximizing the modularity $Q$, while the small communities can not (See the network examples of this type in Section 3). From macro- to micro-scales, i.e., from the whole network as a sole group to the network splitting into a set of single-vertex groups, we can browse the communities at different scales by maximizing $Q^{R\&B}$ with different values of $\gamma$.

In the case of $\gamma \to 0$, which is a lower bound denoted by $\gamma_{min}$, as discussed above, the whole network will be regard as one large group by the optimization of the RB modularity, and no meaningful scales can be found below $\gamma_{min}$.

With increasing the value of $\gamma$ from $\gamma_{min}$, the structures under the network can gradually be revealed by the optimization of the RB modularity. Of course, the community stability is continuously decreasing in the process. When $\gamma =1$, we can obtain the same result as that by optimizing the modularity $Q$. When $\gamma >1$, we can have access to the substructures under those at $\gamma =1$, which are invisible for the modularity $Q$. Especially, the small communities being connected each other, denoted by $s$ and $t$, will become detectable for the RB modularity, when the value of $\gamma$ satisfies

$$\gamma > \gamma_1 = \frac{2M \cdot E_{st}}{k_s k_t}, \qquad (2.3)$$

where the notations are the same as in the inequality (2.2), and $\gamma_1$ is the value of $\gamma$ at the transition point.

Generally, we hope that after the small communities become visible, there should appear a stable community division of network. The stable division is not affected by the increase of $\gamma$ (but the stability of all the communities is gradually decreasing), until the large communities break up. We can deduce the expression of $\gamma$ about the transition point (see Appendix), which is denoted by

$$\gamma_2 \simeq \frac{2M\alpha_l}{k_l} \qquad (2.4)$$

where $M$ the number of edges in the network, and $\alpha_l = k_l^{in}/k_l$ stands for the ratio between the inner degree $k_l^{in}$ and the total degree $k_l$ of the community $l$, which can also be regarded as an indicator of community strength. According to the expression of $\gamma_2$, we can see that the larger the communities, the more quickly they are to break up; that is, the stability of the large communities is easier to be destroyed with the increase of $\gamma$. When $\gamma > \gamma_2$, the large community $l$ will be split by the optimization of the RB modularity.

When the value of the parameter $\gamma$ is large enough, the whole network will be separated into a set of single-vertex groups by maximizing the RB modularity. In this case, the value of $\gamma$ satisfies

$$E_{ij} - \gamma \frac{k_i k_j}{2M} < 0 \quad \text{for all vertices} \quad i \neq j, \qquad (2.5)$$

where $E_{ij} \equiv 1$ in un-weight networks, $k_i$ and $k_j$ are respectively the degrees of vertices $i$ and $j$, and $M$ is the number of edges in network. The smallest value of $\gamma$ that satisfies the inequality, denoted by $\gamma_{max}$, can be evaluated by

$$\gamma_{max} = 2M/k_{min}^2, \qquad (2.6)$$

where $k_{min}$ is the smallest degree of vertices.

By analyzing the RB multi-resolution process, we obtained four expressions of parameter $\gamma$ ($\gamma_{min}$, $\gamma_1$, $\gamma_2$ and $\gamma_{max}$). The first and the last correspond respectively to the beginning and ending of the process, that is, the boundary of the range of the tunable parameter $\gamma$. According to the discussion above, they have no relation to the community structure and are only determined by the degrees of vertices, while the second and the third are not so. Thus the states corresponding to $\gamma_1$ and $\gamma_2$ will be discussed in details.

### 2.3. Analysis of Limitation

Generally, after the small communities become visible, there should exist a stable region ($\gamma_1 < \gamma < \gamma_2$) where all the embedded communities in the network can be detected simultaneously by the optimization of the RB modularity. This is the process we expect, and is to occur when $\gamma_1 < \gamma_2$.

However, it is also possible to occur that the large communities first are broken up by increasing the value of the parameter $\gamma$, before the small communities become detectable for the RB method. This is not that we expected, while it is indeed to happen, when $\gamma_1 > \gamma_2$. According to the expressions of $\gamma_1$ and $\gamma_2$, we can rewrite the inequality as

$$\alpha_l k_s k_t < k_l E_{st} \qquad (2.7)$$

where $k_s$, $k_t$ and $k_l$ are the total degrees of the communities $s$, $t$ and $l$, $E_{st}$ is the total number of edges connecting communities $s$ and $t$, and $\alpha_l = k_l^{in}/k_l$ stands for the ratio between the inner degree and the total degree of the community $l$. For simplicity, suppose that two small communities, having the same total degree $k_s = k_t$, are connected by a single edge ($E_{st} = 1$), and $\alpha_l \simeq 1$ for the community $l$. We can rewrite the inequality (2.7) as $k_s < \sqrt{k_l}$, which indicates a characteristic scale of the resolution where the RB method can not work.

The inequality (2.7) suggests that the RB multi-resolution method in community detection may has some intrinsic limitation—with increasing the value of the tunable parameter, large communities may have been split before small communities become visible in some case. According to the inequality (2.7), the limitation depends on the degree of interconnectedness of small communities and the difference between the sizes of small communities and large communities, while independent of the size of the whole network.

As we see, according to the inequality (2.2), the resolution limit of modularity that is related to the difference between the sizes of small communities and the whole networks can be adjusted by the RB multi-resolution method, but according to the inequality (2.7), we can find that the RB method will encounters another limitation of resolution again that is caused by the difference between the sizes of small communities and large communities in the networks, when the distribution of the sizes of communities is very broad. This is because it is at the cost of the community stability that the enhancement of the modularity resolution is obtained. By increasing the value of the tunable parameter, the stability of all the communities is to be weakened, which makes the communities under threat of splitting, although we can make the small communities gradually disconnected in this process so as to go deep into the substructures in networks. We can infer that all the methods that enhance the modularity resolution at the cost of the community stability may have the limitation similar to the RB method.

### 3. Test

In this section, the network examples mentioned above are realized, and are used to test the theoretical results, especially the condition of the limitation of the RB method. The networks consist of two types of communities with different sizes, and all the communities are cliques (full-connected sub-graphs). The numbers of vertices in large cliques and in small cliques denote respectively by $n_1$ and $n_2$ ($n_1 > n_2$), and the numbers of the large cliques and the small cliques in the networks are respectively $m$ and $2m$. The cliques are connected one by one by single edges, generating a simple ring-like configuration, and there are two small cliques being connected each other between large cliques (see figures 1(a) and 2(a)). Here, we discuss un-weighted and un-directed networks, and the small cliques are invisible for the modularity $Q$. In the following tests, we will scan the community divisions of the networks corresponding to different $\gamma$-values by using the fast greedy algorithm of Blondel et al. [21], and check the experiments by using extremal-optimization algorithm [18].

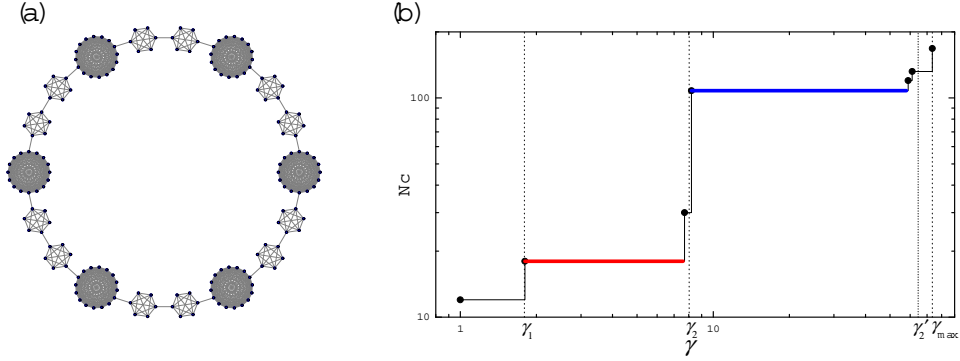

**Figure 1.** (a) Network with $m=6$, $n_1=16$ and $n_2=6$ described in text, where there exists no limitation discussed in text for the RB method. (b) The curve of the value of the parameter $\gamma$ and the number $Nc$ of groups found by maximizing the RB modularity in the network. The bold solid line between $\gamma_1$ and $\gamma_2$ corresponds to that all of the cliques can be revealed. The bold solid line between $\gamma_1$ and $\gamma_2'$ corresponds to that the large cliques break up while the small cliques still is visible.

According to the description above, we firstly generate a network with $m=6$, $n_1=16$ and $n_2=6$ (see figure 1(a)). In this network, we can directly examine the values of the $\gamma_{min}$, $\gamma_1$, $\gamma_2$ and $\gamma_{max}$. It is clear that $\gamma_1 < \gamma_2$, that is, the inequality (2.7) is not satisfied. For the special network consisting of cliques, it means that the small cliques can be detected by the RB method before the large ones break up. Here, we discuss the multi-resolution process for finding the communities under the resolution of modularity $Q$, and focus on the transition states corresponding to $\gamma_1$ and $\gamma_2$. Now, we will analyze the community division of the network obtained by the optimization of the RB modularity with different values of $\gamma$ that is to vary from 1 to $\gamma_{max}$, and compare with the theoretical results in Section 2.

The figure 1(b) shows the curve of the value of $\gamma$ and the number $Nc$ of groups found in the network. As we see, when $1 \leq \gamma < \gamma_1$, the community division of network found has 12 groups, including six large cliques and six groups consisting of small cliques being connected each other, which is the same as that by the optimization of the modularity $Q$. When $\gamma_1 < \gamma < \gamma_2$, all of the cliques become visible, so the number of groups detected is $3m=18$, that is, the number of all cliques in the network (see the bold solid line between $\gamma_1$ and $\gamma_2$ in figure 1(b)).

When $\gamma_2 < \gamma < \gamma_2'$, $\gamma_2'$ being the value of $\gamma$ when the small cliques break up, the large cliques break up while the small cliques still is visible; thus there are 108 groups detected, which are 12 small cliques and 96 vertices from the large cliques (see the bold solid line between $\gamma_1$ and $\gamma_2'$ in figure 1(b)). When $\gamma_2' < \gamma < \gamma_{max}$, the small cliques also will break up. When $\gamma \to \gamma_{max}$, the network splits into a set of single-vertex groups.

As we see in figure 1(b), the test results in the network are consistent with the theoretical analysis, though the values of $\gamma$ when the cliques break up are slightly smaller than the expected values (see Appendix for the reason). Moreover, it is clear that the resolution problem of modularity in the network can be thoroughly solved by the RB method, because the difference between the sizes of small cliques and large cliques is not very large and the limitation discussed in section 2 does not appear in this example.

According to the inequality (2.7), we can easily find the networks where the RB method has the limitation. To show the limitation of the RB method, we create a network with $m=6$, $n_1=48$ and $n_2=6$, according to the description above (see figure 2(a)). Focusing on the values of $\gamma_1$ and $\gamma_2$, we can find $\gamma_1 > \gamma_2$, i.e., the inequality (2.7) is satisfied, so the large cliques are to split before the small cliques become visible for the RB method.

Similar to the above test, we also analyze the community division of the network obtained by the optimization of the RB modularity with different values of $\gamma$ varying from 1 to $\gamma_{max}$. The figure 2(b) displays the curve of the value of $\gamma$ and the number $Nc$ of groups found in the network. As shown in figure 2(b), when $1 \leq \gamma < \gamma_2$, the community division of network found consists of 6 large cliques and 6 groups formed by the mergers of small cliques. When $\gamma_2 < \gamma < \gamma_1$, the large cliques have broken up, while the small cliques are still invisible for the RB modularity. When $\gamma_1 < \gamma < \gamma'_2$, $\gamma'_2$ being the value of $\gamma$ when the small cliques break up, the small cliques are visible for the RB modularity, while the large cliques have been split into individual vertices, so there are 300 groups detected, which are 12 small cliques and 288 vertices from the large cliques (see the bold solid line between $\gamma_1$ and $\gamma'_2$ in figure 2(b)). When $\gamma'_2 < \gamma < \gamma_{max}$, the small cliques also begin to break up. When $\gamma \to \gamma_{max}$, the network split into a set of single-vertex groups.

Clearly, the test results are also consistent with the theoretical analysis. More importantly, it clearly indicates the existence of the limitation discussed above for the RB method in this test network. The RB method can not deal with the limitation problem caused by the difference between the sizes of small communities and of large communities in the networks, while it can finally find all small cliques in the network, that is, can solve the resolution limit of modularity that is related to the difference between the sizes of small communities and the whole networks.

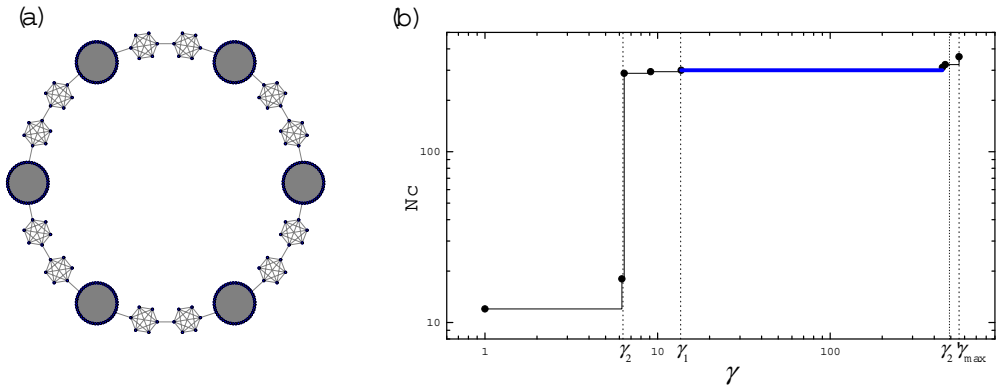

**Figure 2.** (a) Network with $m = 6$, $n_1 = 48$ and $n_2 = 6$ described in text, where the limitation of the RB method appears. (b) The curve of the value of the parameter $\gamma$ and the number *Nc* of groups found by maximizing the RB modularity in the network. The bold solid line between $\gamma_1$ and $\gamma'_2$ corresponds to that the small cliques are visible for the RB method while the large cliques have been split into individual vertices.

Finally, we analyze the toy model network proposed by Arenas et al., where the RB method is invalid [32]. The network consists of a clique and four star-like structures of different sizes. The clique in the network of Arenas et al. has 10 vertices. In figure 3(a), we show the structure of the network of this type, while it is the only difference that the clique here contains 7 vertices. In the toy model network of Arenas et al. we can find that the inequality (2.7) is satisfied (i.e. $\gamma_1 > \gamma_2$, where $\gamma_1$ and $\gamma_2$ are respectively the values of $\gamma$ when the smallest community becomes visible and when the clique begins to break up). Clearly, the limitation of the RB method appears in the network—the clique will break up before the small communities become visible for the RB modularity. This is the reason that the RB method is invalid in this network.

Now, we can reduce the size of the clique in the network to escape from the limitation of the RB method. For example, the figure 3(b) shows that the RB method can find the all communities in the new network where there are only 7 vertices in the clique.

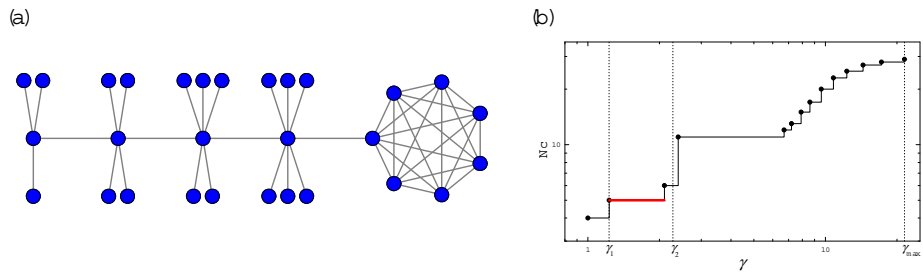

**Figure 3**. (a) A toy model network consisting of a clique and four stars of different sizes. (b) The curve of the value of the parameter $\gamma$ and the number *Nc*

of groups found by maximizing the RB modularity, in the toy model network with communities of different sizes and densities.

## 4. Discussion

We have discussed the limitation of the RB method in community detection, while whether the AFG multi-resolution method also has the limitation? The multi-resolution method proposed by Arenas et al. (AFG) is to provide each vertex with a self-loop of the same magnitude $r$, which is equivalent to modifying the modularity function $Q$ by the parameter $r$. The modified modularity function of the AFG method [32] can be written as

$$Q^{AFG}(r) = \sum_s \left( \frac{k_s^{in} + n_s r}{2M + Nr} - (\frac{k_s + n_s r}{2M + Nr})^2 \right) \quad (4.1)$$

where $r$ is the tunable parameter, $n_s$ is the number of vertices in community $s$, $k_s^{in}$ is the inner degree within the group $s$, $k_s$ is the total degree of group $s$, and $N$ is the number of vertices in the network. According the modularity function, we can find that the community stability is also to degenerate with the increase of the parameter $r$.

It has been shown that the AFG method can adjust the resolution of the modularity to analyze the communities at different scales, by tuning its parameter [32]. Similar to the RB method, the result obtained by optimizing the modularity $Q$ in Eq. (1.1) corresponds to $r=0$. When $r<0$, we can find the superstructures above those at $r=0$. When $r>0$, we can obtain the substructures under those at $r=0$. Generally, the results of the AFG multi-resolution procedure differ from the RB, but they can give the same results respectively at the beginning and ending of their procedures: the whole network is regarded as a group and the whole network will be separated into a set of single-vertex groups. The comparison of these methods has been discussed before [32-33]. Here, we discuss the limitation of the AFG method, similar to that of the RB method.

It is difficult to give general formulas similar to those in the RB method, because of the special way by which the AFG modifies the modularity $Q$. But, interestingly, the AFG method is equivalent to the RB method for all divisions in the homogeneous networks where all of vertices have the same degree. In this case, if we denote the degree of every vertex by $\bar{k}$ and replace the parameter $r$ by $r'\bar{k}$, and then the relation between the tunable parameters of two methods can be expressed as (see ref [32] for the relation)

$$\gamma = \frac{2M}{2M + Nr} \cdot \left( \frac{\bar{k} + r}{\bar{k}} \right)^2 = 1 + r'. \quad (4.2)$$

By using the multi-resolution procedures of the RB and AFG methods, the same results will be obtained, and therefore, the limitation in the RB method is also to occur in the AFG method for the homogeneous networks.

Moreover, we believe that the inequality (2.7) being satisfied suggests the presence of the limitation of the multi-resolution methods, but the inequality (2.7) being not satisfied does not always means the disappearance of the limitation. Because the expression of $\gamma_2$ is obtained based on ensemble average, while the splitting of the random communities may be much easier than that expected, due to the fluctuations in random graphs (see Appendix and ref [35-36]). These communities may break up before the value of $\gamma$ approaches to the expected value of $\gamma_2$ in the multi-resolution methods. We can infer that the limitation of the multi-resolution methods discussed above may come forth in general networks before the inequality (2.7) is satisfied. Therefore, the inequality (2.7) may be regarded as sufficient condition that the limitation of the multi-resolution methods appears in network, while not necessary condition.

To clearly show the above conjecture as well as the limitation of the AFG method, we generate a homogeneous network of the ring-like structure with $m=6$, $n_1=24$ and $n_2=6$, by retaining the community structure of the network in Section 3 while making all the vertices in the network have the same degree of 6 (see figure 4(a)). In this network, the small communities are still cliques (full-connected sub-graphs), and the small cliques nearby are connected by five edges; while the large communities are random sub-graphs. The RB and AFG methods will give the same results in this network.

Examining the inequality (2.7), we can find that it is not satisfied in the network (i.e. $\gamma_1 < \gamma_2$, where $\gamma_1$ and $\gamma_2$ are respectively the values of $\gamma$ when the small cliques become visible, and when the large communities begin to break up). However, we find that the large communities have broken up in the network before the small communities become visible by using the multi-resolution methods (see figure 4(b)). This result clearly show that both the RB and AFG methods have the limitation problem discussed above in the network, and it indeed comes forth in advance before the inequality (2.7) is satisfied, due to the random fluctuations in communities.

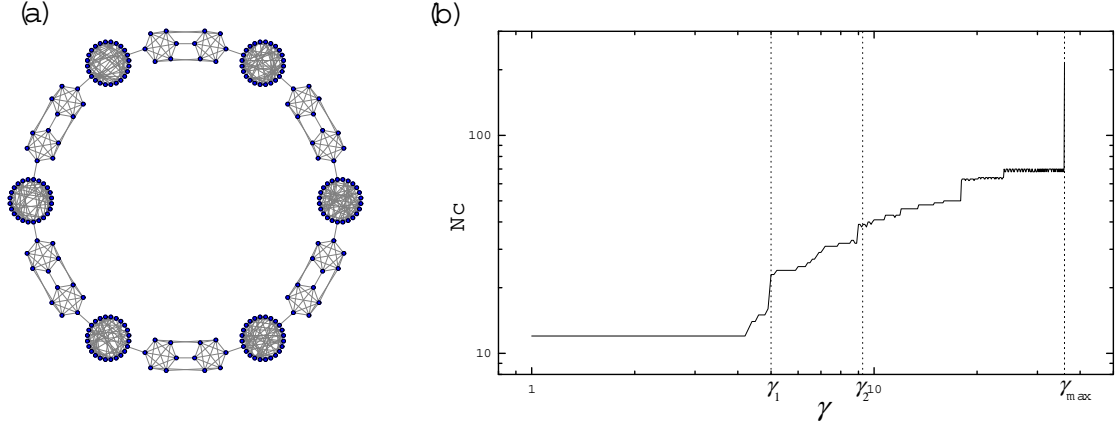

**Figure 4.** (a) Network with $m = 6$, $n_1 = 24$ and $n_2 = 6$ described in text, which is a variant of the clique-loop network in section 3. (b) The curve of the value of the parameter $\gamma$ and the number *Nc* of groups found by using the RB and AFG method in the network.

## 5. Conclusion

In this paper, we presented a critical analysis of the applicability of the (RB and AFG) multi-resolution methods to the problem of community detection. We show that both the methods have the intrinsic limitation —large communities may have been split before small communities become visible, even in the extreme case where communities are full-completed sub-graphs. This is because it is at the cost of the community stability that the enhancement of the modularity resolution is obtained. By increasing the value of the tunable parameters, the stability of all the communities is to be weakened, which makes the communities under threat of splitting, though we can make small communities gradually disconnected in this process. Moreover, we can infer that the methods that enhance the modularity resolution at the cost of the community stability may encounter the similar limitation.

The theoretical analysis and the experimental tests in several network examples indicated that the limitation depends on the degree of interconnectedness of small communities and the difference between the sizes of small communities and large communities, while independent of the size of the whole network. That is to say, the multi-resolution methods can not deal with the resolution problem caused by the difference between the sizes of small communities and of large communities in the networks, though it can finally find out all small pre-defined communities in the networks by the increase of their resolution parameters, that is, can solve the resolution limit of modularity that is related to the difference between the sizes of small communities and the whole networks.

It is worth noticing that, the splitting of communities in network may be much easier that that expected, due to the random fluctuations [35-36]. The limitation of the multi-resolution methods may in advance appear in general networks, that is, coming forth before the inequality (2.7) is satisfied. Therefore, the inequality (2.7) may be regarded as sufficient condition that the limitation of the multi-resolution methods appears in network, while not necessary condition.

Moreover, we find that A. Lancichinetti and S. Fortunato [37] have a discussion of interest more recently that is closely related to the multi-resolution methods' limitation similar to that in this paper, and G. Krings and V. D. Blondel [38] also discussed the problem of large and dense communities being to disaggregate with the increase of the resolution parameter. As we know, these multi-resolution methods indeed can help us find the communities of networks at different scales, to some extent, by varying their parameters to adjust the resolution of the modularity. But these methods themselves, before being applied to the real problem of community detection, should have been thoroughly understood to ensure that the multi-scale structures found in networks are reasonable. So it should be worthwhile paying attention to the limitation of these multi-resolution methods, especially when the distribution of the sizes of communities is very broad. We hope that the study in the paper can help us further understand the applicability of the multi-resolution methods in community detection.

**Appendix**

Here, we give the mathematical proofs of the expression of the parameter $\gamma$ in the case that the communities break up by using the RB method. Given a community $l$, the total degree and the inner degree of it are denoted respectively by $k_l$ and $k_l^{in}$. We define the ratio between $k_l^{in}$ and $k_l$ by $\alpha_l = k_l^{in}/k_l$, which can be regarded as an indicator of community strength. The number of vertices in the community is $n_l$.

Now, we choose a random division of the community into two parts $a$ and $b$, where the degrees of the two parts are denoted respectively by $k_a$ and $k_b$ ($k_l = k_a + k_b$). With increasing the value of the parameter $\gamma$, the community $l$ will break up by using the RB method, when

$$E_{ab} - \gamma \frac{k_a k_b}{2M} < 0, \text{ or } \gamma > \frac{2M \cdot E_{ab}}{k_a k_b}, \quad \text{A.1}$$

where $M$ is the total number of edges in the network, $E_{ab}$ is the total number of edges between the parts $a$ and $b$.

To estimate the expression of $\gamma$ at the transition point above, we suppose that there exists the relation $\alpha_i \approx \alpha_l$ for all vertices in the community $l$, where $\alpha_i = k_i^{in}/k_i$ is the ratio between the inner degree $k_i^{in}$ and the degree $k_i$ of vertex $i$ in the community. Statistically, we can obtain the expected value of $E_{ab}$, $E_{ab}^{expect} = \alpha_l k_a k_b / (k_l - \bar{k})$, where $\bar{k}$ is the average degree of vertices in the community and $k_l$ minus $\bar{k}$ is to exclude the effect of self-links. Whereupon we can rewrite (A.1) as

$$\gamma > \gamma_2 = \frac{2M \alpha_l}{k_l} \cdot \frac{n_l}{n_l - 1} \simeq \frac{2M \alpha_l}{k_l}, \text{ when } n_l \text{ is very large.} \quad \text{A.2}$$

If the inequality is satisfied, then the community $l$ will break up or have broken up.

Notice that the inequality may be regarded as a sufficient condition that the community breaks up, while not necessary, due to the above hypothesis and random fluctuations in communities [35-36]. There are at least the reasons in two aspects.

For convenience, we supposed that $\alpha_i \approx \alpha_l$ for all vertices in the community $l$, however it is also possible that there exists $\alpha_i < \alpha_l$ in some parts of the community. These parts are less stable than those with $\alpha_i \geq \alpha_l$, so they may have been separated from the community before the inequality is satisfied. This occurs in all test networks of the paper.

Moreover, the expression of $E_{ab}^{expect}$ based on ensemble average gives only the expected value of the number of edges between the parts $a$ and $b$. However the number of the possible divisions of the community into two parts is very large, there must exist the divisions whose $E_{ab}$ is much smaller than the expected value [27, 35-36], especially in large random sub-graphs. So the community $l$ may have broken up before the value of $\gamma$ approaches to $\gamma_2$, in random networks with community structure. In other words, the splitting of random communities will be much easier than that expected by the inequality (A.2), especially in large communities, while the cliques will be exceptions.